\documentclass{webofc}

\usepackage[varg]{txfonts}   
\usepackage{hyperref}
\usepackage{url}
\hypersetup{colorlinks=true,citecolor=blue,urlcolor=blue,linkcolor=blue}

\newcommand{\vb}{\vec}
\renewcommand{\vec}[1]{\mathrm{\mathbf{#1}}}
\newcommand{\dd}[2][]{\mathrm d^{#1}{#2}\,} 
\newcommand{\dv}[2][]{\frac{\dd{#1}}{\dd{#2}}}
\newcommand{\pdv}[2][]{\frac{\partial{#1}}{\partial{#2}}}

\newcommand{\Conetwo}{\mathcal {C}^{1\leftrightarrow 2}}
\newcommand{\Ctwotwo}{\mathcal{ C}^{2\leftrightarrow 2}}

\begin{document}
\title{Pre-hydrodynamic jet momentum broadening beyond the jet quenching parameter}

\begin{flushright}
MIT-CTP/5933
\end{flushright}

\author{\firstname{Alois} \lastname{Altenburger}\inst{1}\fnsep
\and
        \firstname{Kirill} \lastname{Boguslavski}\inst{1,2}\fnsep
         \and
        \firstname{Florian} \lastname{Lindenbauer}\inst{1,3}\fnsep\thanks{Speaker, \email{flindenb@mit.edu}\\
        FL is a recipient of a DOC Fellowship of the Austrian Academy of Sciences at TU Wien (project 27203).
        This work is funded in part by the Austrian Science Fund (FWF) under Grant DOI 10.55776/P34455, and Grant DOI 10.55776/J4902 (FL).
        For the purpose of open access, the authors have applied a CC BY public copyright
        license to any Author Accepted Manuscript (AAM) version arising from this submission.
        The results in this paper have been achieved in part using the Austrian Scientific Computing (ASC) infrastructure, project 71444.
        }
}

\institute{Institute for Theoretical Physics, TU Wien, Wiedner Hauptstraße 8-10, 1040 Vienna, Austria 
\and
           SUBATECH UMR 6457 (IMT Atlantique, Université de Nantes,
           IN2P3/CNRS), 4 rue Alfred Kastler, 44307 Nantes, France 
\and
           MIT Center for Theoretical Physics -- a Leinweber Institute,
           Massachusetts Institute of Technology, Cambridge, MA 02139, USA
          }

\abstract{
	We obtain the collision kernel and related dipole cross section during the initial nonequilibrium stages in heavy-ion collisions. These quantities are a crucial input for jet quenching calculations. We further compute the gluon splitting rates in the AMY formalism resulting from this nonequilibrium kernel. Comparing with thermal and commonly used forms, we find that particularly the gluon splitting rate for parton energies of the order of the hard effective temperature significantly differs from these approximations.
}
\maketitle
\section{Introduction}
\label{intro}
Jets in heavy-ion collisions may reveal properties of the strong interaction. They originate from energetic partons created in the initial collision and lose energy when traversing the created quark-gluon plasma. A crucial input for jet energy loss calculations is the dipole cross section $C(\vb b)$ \cite{Apolinario:2022vzg}. Its small distance form is determined by the jet quenching parameter $\hat q$, which physically encodes transverse momentum broadening,
\begin{align}
	C(\vb b) = \int\frac{\dd[2]{\vb q_\perp}}{(2\pi)^2}\left(1-e^{i\vb q_\perp\cdot \vb b}\right)C(\vb q_\perp),\label{eq:fouriertrafo} && C(\vb b)\approx \frac{1}{4}\hat q\vb b^2, &&     \hat q = \dv[\langle p_\perp^2\rangle]{t}=\int\frac{\dd[2]{\vb q_\perp}}{(2\pi)^2}\vb q_\perp^2 C(\vb q_\perp).
\end{align}
Recent increased interest in understanding jet-medium interactions during the nonequilibrium initial stages in heavy-ion collisions has led to the computation of $\hat q$ during the early stages in heavy-ion collisions \cite{Ipp:2020nfu, Carrington:2021dvw, Avramescu:2023qvv, Boguslavski:2024ezg, Boguslavski:2024jwr}. Under the assumption that the system is sufficiently weakly coupled, the earliest instants directly after the collision are dominated by large overoccupied classical gluon fields, a state called the \emph{Glasma}. After about $\tau\sim 1/Q_s$, where $Q_s$ is the saturation momentum, the system admits a description in terms of gluons, whose distribution function $f(\vb p,\tau)$ follows an effective kinetic Boltzmann equation \cite{Arnold:2002zm},
\begin{align}
	\pdv[f(\vb p)]{\tau}- \frac{p_z}{\tau} \pdv[f(\vb p)]{p_z} =-\Conetwo[f(\vb p)]- \Ctwotwo[f(\vb p)].
	\label{eq:boltzmann_equation}
\end{align} 
This QCD effective kinetic theory can be used to connect the system to the later hydrodynamic stage \cite{Kurkela:2015qoa, Kurkela:2018wud}.
Here, going beyond $\hat q$, we discuss how we obtain the collision kernel $C(\vb q_\perp)$, the dipole cross section $C(\vb b)$ and the gluon splitting rates $\gamma$ \cite{Arnold:2002ja} using QCD kinetic theory simulations \cite{kurkela_2023_10409474} with the isoHTL screening prescription \cite{Boguslavski:2024kbd}. We provide more details in Ref.~\cite{Altenburger:2025iqa}.

\section{Obtaining the collision kernel}
\label{sec-1}
\begin{figure}
	\centering
	\centerline{
		\includegraphics[width=0.45\linewidth]{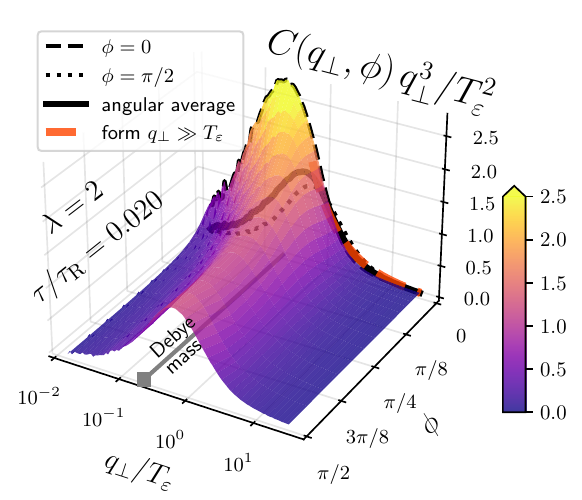}
		\includegraphics[width=0.45\linewidth]{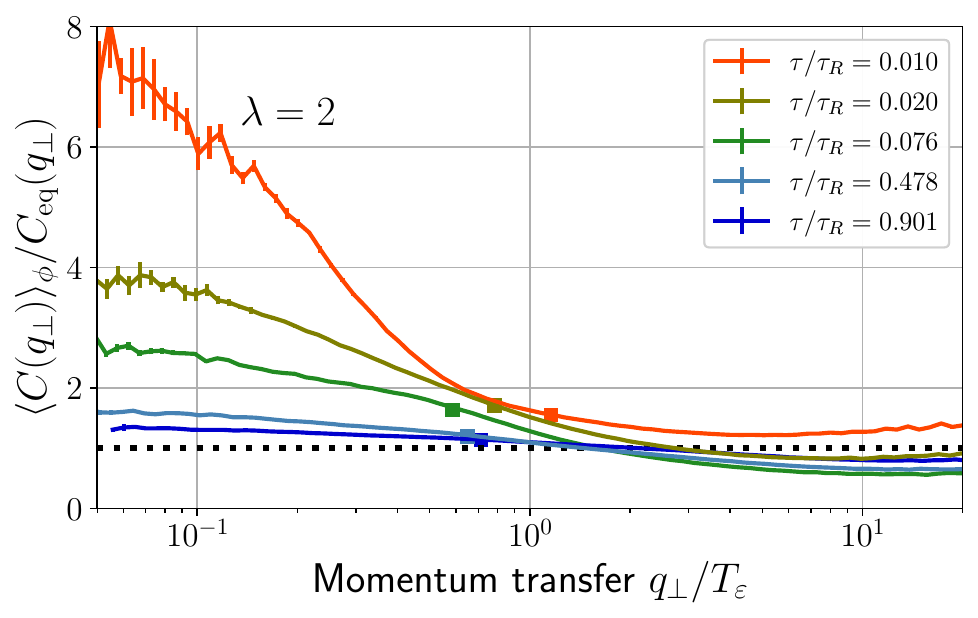}
	}
	\caption{Results for the collision kernel in momentum space. (Left): For a single time as a function of transverse momentum and angle. (Right): Angular averaged collision kernel as a function of transferred momentum for different times (color-coded). Figures from \cite{Altenburger:2025iqa}.}
	\label{fig:collkern}       
\end{figure}
The collision kernel $C(\vb q_\perp)$ can be obtained similarly to the jet quenching parameter 
\begin{align}
	\hat q^{ii}=\int_{\substack{q_\perp <\Lambda_\perp\\ p\to \infty}}\dd{\Gamma}\left(q^i\right)^2\left|\mathcal M\right|^2 f(\vb k)\left(1+f(\vb k')\right), &&C(\vb q_\perp)=\int_{\substack{p\to \infty}}\dd{\tilde \Gamma}\left|\mathcal M\right|^2 f(\vb k)\left(1+f(\vb k')\right),
	\label{eq:formula-qhat-collisionkernel}
\end{align}
where $|\mathcal M|^2$ is the isoHTL screened matrix element \cite{Boguslavski:2024kbd}. We present the results in Fig.~\ref{fig:collkern}, where we show the collision kernel for a fixed time in the left, and its angular average for various times (color-coded) in the right panel, highlighting the differences to equilibrium. 
The left panel depicts the contribution to $\hat q$, which in equilibrium is peaked at the Debye screening mass. We observe that for early times, the position of this peak becomes angular-dependent, suggesting an angular-dependent effective mass.
In the right panel, we compare to the corresponding (Landau-matched) equilibrium system at the same energy density and observe that small-momentum exchange is significantly enhanced, particularly at early times.

\section{Dipole cross section and gluon splitting rates}
\begin{figure}
	\centering
	\centerline{
		\includegraphics[width=0.45\linewidth]{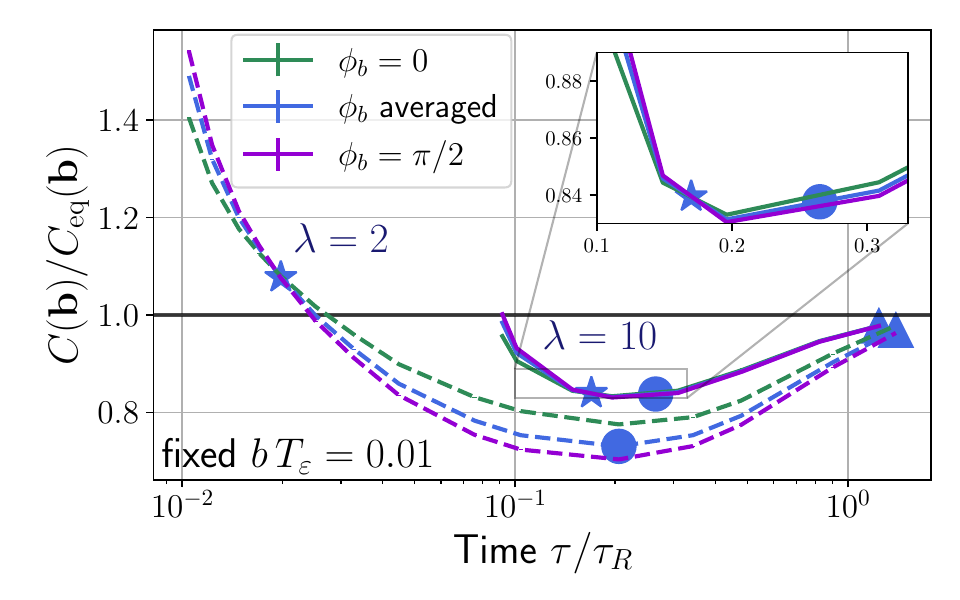}
		\includegraphics[width=0.45\linewidth]{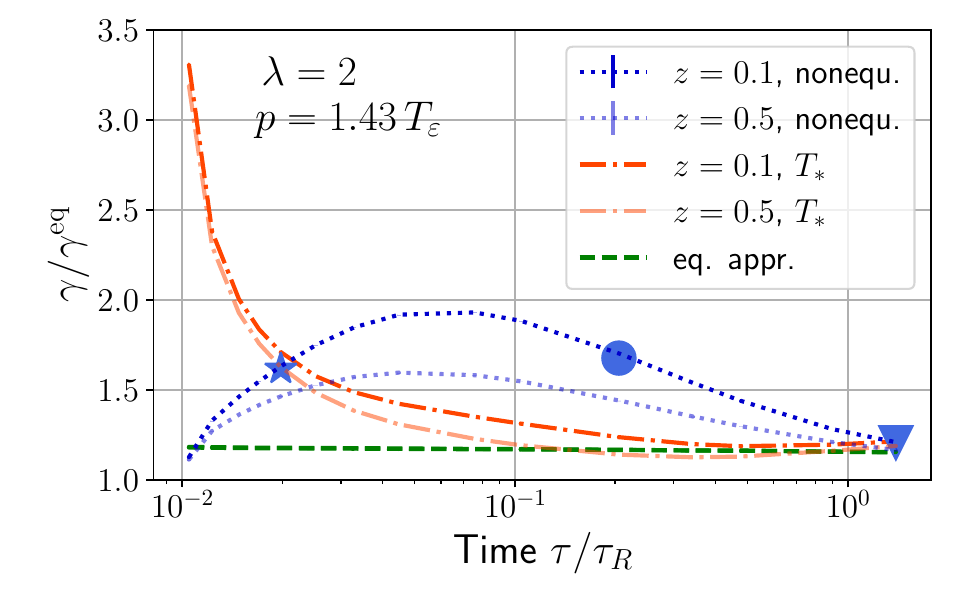}
	}
	\caption{(Left): Dipole cross section for fixed $b$ as a function of time. (Right): Gluon splitting rate for a fixed parton energy $p$ as a function of time, normalized to equilibrium. The curve obtained from the nonequilibrium collision kernel is depicted as a blue dotted line. The orange dash-dotted curve depicts the rate using a common approximation. Figures from \cite{Altenburger:2025iqa}.}
	\label{fig:rates}       
\end{figure}
We now move on to compute the dipole cross section $C(\vb b)$ using Eq.~\eqref{eq:fouriertrafo} and show the result for a fixed small $b=0.01/T_\varepsilonup$ in the left panel of Fig.~\ref{fig:rates}. For $\lambda=2$, it is initially above the thermal value, but then drops below and approaches the thermal value from below. For the phenomenologically more relevant coupling $\lambda=10$ (solid lines), the dipole cross section is reduced throughout the whole time evolution.

We now move on to use the full form of the dipole cross section as input to calculate the gluon splitting rates in the AMY formalism \cite{Arnold:2002ja},
\begin{align}
    \gamma&=\frac{p^4+p'^4+k'^4}{p^3p'^3k'^3}\frac{d_A\alpha_s}{2(2\pi)^3}\int \frac{\dd[2]{\vb h}}{(2\pi)^2}2\vb h\cdot \mathrm{Re} \vb F. \label{eq:gammarate} \\
	2\vb h&=i\delta E(\vb h)\vb F(\vb h)+\frac{g^2}{2}\int\frac{\dd[2]{\vb q_\perp}}{(2\pi)^2}C(\vb q_\perp)[(3\vb F(\vb h)-\vb F(\vb h-p\vb q_\perp)-F(\vb h-k\vb q_\perp)-\vb F(\vb h+p'\vb q_\perp))]\label{eq:integral-eq}
\end{align}
with $\delta E(\vb h)=m_D^2/4\times (1/k+1/p-1/p')+h^2/(2pkp')$, and the underlying assumptions of an infinite medium and static collision kernel during the splitting process. Despite these assumptions and approximations, these rate equations are used in QCD kinetic theory simulations.

We solve the integral equation \eqref{eq:integral-eq} with the anisotropic collision kernel $C(\vb q_\perp)$ using a novel numerical method \cite{Lindenbauer:2025ctw}, and depict the result for fixed parton energy as a blue dotted curve in Fig.~\ref{fig:rates}. The curve is normalized to the equilibrium rate. We compare with a commonly used approximation, denoted $T_*$ in the figure, which uses an isotropic approximation of the collision kernel when solving the integral equation \eqref{eq:integral-eq}. Fig.~\ref{fig:rates} reveals that this approximation leads to a strongly qualitatively and quantitatively different evolution of the rate, with yet unexplored consequences for QCD kinetic theory simulations.

\section{Conclusion}
We computed the probability $C(\vb q_\perp)$ of a jet parton to receive a momentum kick $\vb q_\perp$, and calculated the dipole cross section $C(\vb b)$, related via Eq.~\eqref{eq:fouriertrafo}. This is a crucial ingredient for jet energy loss calculations. Furthermore, using common assumptions, we obtained the gluon splitting rate from the nonequilibrium kernel. We contrast and compare our results to calculations in thermal equilibrium, which are commonly employed in place of a full nonequilibrium formalism. This may pave the way for a more realistic modeling of jets during the initial nonequilibrium stages in heavy-ion collisions and improve QCD kinetic theory simulations to model the early nonequilibrium plasma.

\bibliography{bib}

\end{document}